# The start of the Sagittarius spiral arm (Sagittarius origin) and the start of the Norma spiral arm (Norma origin) - model-computed and observed arm tangents at galactic longitudes  -20° <  l < +23°

Jacques P. Vallée

National Research Council Canada,  National Science Infrastructure, Herzberg Astrophysics, 5071 West Saanich Road, Victoria, B.C., Canada V9E 2E7

## Abstract

Here, we fitted a 4-arm spiral model to the more accurate data on global arm pitch angle and arm longitude tangents, to get the start of each spiral arm near the galactic nucleus. We find that the tangent to the 'start of the Sagittarius' spiral arm (arm middle) is at l= -17° ±0.5°, while the tangent to the 'start of the Norma' spiral arm (arm middle) is at l= +20° ±0.5°.

Earlier, we published a compilation of observations and analysis of the tangent to each spiral arm tracer, from longitudes +23° to +340°; here we cover the arm tracers in the remaining longitudes  +340 (= -20°)  to +23°.  Our model arm tangents are confirmed through the recent observed masers data (at the arm's inner edge).  Observed arm tracers in the inner Galaxy show an offset from the mid arm;  this was also found elsewhere in the Milky Way disk (Vallée 2014c).

In addition, we collated the observed  tangents to the so-called  '3-kpc-arm' features; here they are found statistically to be near l = -18° ± 2° and near l = +21° ±2° , after excluding misidentified spiral arms.    We find that the model-computed arm tangents in the inner Galaxy are  spatially coincident with the mean longitude of the observed  tangents to the '3-kpc-arm' features  (same galactic longitudes, within the errors).  These spatial similarities may be suggestive of a contiguous space.

## 1.  Introduction

The spiral arms in the Milky Way's disk extends out to perhaps 15 kpc.  Being in the disk itself, we can see an increased intensity from tracers at certain galactic longitudes, when we look through a spiral arm tangentially. Arm tracers include stars, HII regions,  masers, etc. A catalog of such tracers, and the longitudes tangent to each spiral arm, was given by   Vallée (2014c), covering  the arm tracers from longitudes +23° to +340°.  Arms can also contain  gas (atomic, molecular, ionized), dust (hot and cold), and  large-scale magnetic field – for a recent review, see Vallée (2011; 2012), Van Eck et al (2015).  Here  we cover the spiral arms in the remaining longitudes  -20° (=  340°)  to +23°.

Our objective  is to accurately model spiral arms (see equations in Vallée 1995),  to fit observations of arm tangents between the Sun and the Galactic Centre (=GC), in a galactic radial range from 4 to 8 kpc  (arm tangents in Vallée 2014c), using the best observed spiral pitch angle value (see pitch in Vallée 2015), and to predict  the start of  these arms near the GC (arm origins between 2 and  3 kpc).  Predictions are compared to published observed arm tracers near the GC.

The observed 'start of  Perseus' arm, or Perseus arm origin,  is already well documented, between galactic longitudes  -23° (= 337°) for the $^{12}$CO 1-0 peak intensity and -20° (=340°) for the hot dust lane  (Table 2 in Vallée 2008;  Table 3 in Vallée 2014a;  Table 3 in Vallée 2014b;

Table 1 in Vallée 2014c).This arm origin is just outside the zone being investigated here. Similarly, the observed beginning of the Scutum arm is well documented, between galactic longitudes +26$^o$ (dust lane and methanol masers) and +33$^o$ for the $^{12}$CO 1-0 peak intensity (Table 1 in Vallée 2014c). This arm origin is also just outside the zone being investigated here.

Here we seek the origin of the other two spiral arms: Sagittarius and Norma.

Recent published global arm positions near the Sun are listed in Table 1 and examined in Section 2. In Section 3, we ascertain the arm pitch angle over distance (fig.1) and we use the tangents to the spiral arms closer to the sun and the statistical means of the pitch angle for the arms, in order to compute the starts of the arms (fig. 2) in the central galactic longitudes ( $-20^o < l < +23^o$). In Section 4 we collate the shape of the bulge and the nuclear bar (Table 2), the long and short bars (Table 3), and the blurred molecular ring (fig.3). In Section 5 we combine the model with these large observed objects (bars, ring, bulge) in the inner galaxy (fig.4), as well as with published theories say. In Section 6 we collect observed arm tracer data in the inner galaxy (Table 4), searching for tracer offsets.

In addition, in Section 7 we collate data on recently published so-called '3-kpc-arm' features (Table 5; fig. 5). In Section 8 we examine any overlap between these '3-kpc-arm' features and the arm origins , as well as what some published theories say (fig.6). In Section 9, we conclude that the galactic tangent to the start of the Sagittarius arms (in CO) at l= -17$^o$ and that the galactic tangent of the Norma arm at l= +20 (in CO) have been located, and that they are also tangents to the '3-kpc-arm' features there within the observational errors. An interaction of the arms with the bars is possible.

## 2. Positional method – global spiral arm fits

Recent papers on the Milky Way included a map of the disk, with spiral arms sketched onto it. Data from this sketch on the arms can include its shape, its pitch angle, its interarm at the Sun, its number, etc.

In a series of papers, we have catalogued the published observational results since 1980 for the Milky Way's arms (number of arms, arm shape, pitch angle, interarm distance through the Sun's location). Results were put in blocks, each block with a minimum of 15 and a maximum of 20 results. Papers in this series were: Vallée (1995 – Paper I), Vallée (2002 – Paper II), Vallée (2005 – Paper III), Vallée (2008 – Paper IV), Vallée (2013 – Paper V), Vallée (2014a – Paper VI), Vallée (2014b – Paper VII), Vallée (2015 – Paper VIII).

In **Table 1**, we catalogue the most recent published papers on this topic, using the same maximum and minimum number of results per bloc. Fitted data (pitch angle, interarm at the Sun's position, arm shape and number) are taken from published figures employing a global model (their own or one they adopted) with their arm tracers. The last rows showed the most recent statistics; the results do not diverge significantly from the previously published ones (Paper VIII). Some comments are listed in the **Appendix A**, when appropriate.

Over the same time span, some papers gave observational results concerning the precise galactic longitude of the tangent to a spiral arm; these arm tangents in galactic longitude have been collected and statistically analyzed for each arm tracer (CO, dust, O stars, etc), and they yielded for a typical arm the mean offset of a tracer away from the dust lane (Paper VI; Vallée 2014c).

A real spiral arm, when seen tangentially from the Sun along the line of sight, (a) would extent several kpc inside (one narrow tube), and (b) would encompass many different objects

(dust, gas, masers, stars, etc) emitting at different wavelengths (different tracers), and (c) would show an ordered offset among its tracers towards the Galactic Center: dust at the inner arm edge, cold broad CO at its center, etc (ordered offsets), and (d) would have a continuous radial velocity range (no gap in radial velocity for gas belonging to that arm).

Recognized spiral arms elsewhere (Carina, Crux-Centaurus, Norma, Scutum, Sagittarius) were seen in <u>at least 6 different tracers</u> (see Table 1 in Vallée 2014c), and with ordered tracers offset and *reversing* across the longitude l=0$^o$ (dust lane inward toward the GC).

## 3. Newest 4-arm model fit

Here we seek to fine-tune the spiral arm parameters, as seen from the Sun, and thus as they are located between 4 and 8 kpc from the GC. We need to ascertain if we can use this fit in order to extrapolate it downward toward the GC.

3.1 M51 and other nearby disk galaxies

Defining the pitch angle as angle of deviation of the arm tangent from a tangent to an orbital circle, then the pitch angle may deviate from 0 (circle) down to 90$^o$ inward, or up to 90$^o$ outward (a range of 180$^o$).

For several nearby disk galaxies, their pitch appears to be stable with galactic radius on large scale (10 kpc), while some local deviations occur up and down the mean pitch.

In **Figure 1**, we show the observed arm pitch angle values for the spiral galaxy M51 at 9.7 Mpc from us. We used the locations of CO peaks (open triangles for arm 1, filled triangles for arm 2) from Patrickeyev et al (2005 – fig.2), with the CO arms 1 and 2 shown in Fig. 3a of Patrikeev et al (2006). We also superimposed the locations of HII regions (open circles for arm B, filled circles for arm A) from Honig & Reid (2015-tab.5).

The y-axis shows that for M51 most of the full range in pitch angle is empty, save for the narrow range from about 0$^o$ up to about 30$^o$. The circles show a mean pitch angle (black horizontal bar), but with local deviations with an amplitude of about 20$^o$.

In this figure, the CO peak positions were rectified to a face-on position assuming a galactic inclination of 20$^o$ (Fig. 3 in Patrikeev et al 2006), so the orbit of a CO cloud around the galactic center will appear as a circle on a face-on map. Their mean pitch value is near 25$^o$ (dotted black horizontal line). Using optical i-band image, Hu et al (2013 – fig. 1) found an inclination of 20$^o$ and a mean pitch angle near 17$^o$.

The HII region positions (Fig.8 in Honig & Reid 2015) were not deprojected for the galaxy's inclination to the line of sight, so the orbit of an HII region will appear as an ellipse there (axial ratio of 0.94), so the (projected) pitch angle data were extracted with a mean pitch near 12$^o$. We have not deprojected it here, so our figure 1 matches the data in their paper. Deprojection would require to increase some pitch angle values near the minor axis of the ellipse.

Similarly, Honig & Reid (2015 – fig. 9) showed the pitch angle data for several other spiral galaxies (M74, NGC 1232, NGC 3184), concluding that there is no evidence for a general change in pitch angle with galactocentric radius, using the distribution of HII regions in the galactic disk as seen in the optical domain. In M51, the CO data does show a scatter over 2 kpc, from a small upward trend, then a jump down, then another small upward trend over another 2kpc. The overall general trend over 5 kpc is thus rather flat.

Honig and Reid (2015 – sect. 5.1 and fig. 9) stated that "Interestingly, we find no evidence for a general change in pitch angle with galactocentric distance", referring to 4 nearby late-type spiral galaxies (their Fig. 9). They further stated that "Currently the parallax data for the Milky Way typically trace only arm segments about 5 to 10 kpc in length, which corresponds to the scale over which we find relatively constant pitch angles in the external galaxies". They also acknowledged a significant scatter in pitch angle, perhaps with at most a very small systematic trend (smaller than their mean value over 10 kpc).

The pitch angle in the Milky Way is not strictly constant. There is a significant spread around some mean pitch angle value. At a given place in a single arm, we do not know how much the value of the spread is locally, so here we ignore the spread it in our general modelling. We now made it clear that there still could be a very small trend over 8 kpc, unbeknown to us, yet possibly hidden in a local spread.

### 3.2 Milky Way disk galaxy

Thus here we will reasonably assume that the pitch angle does not change appreciably from a galactic distance from 3 to 10 kpc in our Galaxy, save for local effects. In the complex models of Bissantz et al 2003, those with a bar with a different pattern speed than that of the spiral arms, the gas in the arms near the bar stay in the arms (irrespective of the gas in the bar). But when the arms and bar have the same pattern speed, the arms dissolve, so to observe the arms there as we do is because the arms have a different pattern speed than the bar. Bissantz et al preferred 60 $Gyr^{-1}$ for the bar, and 20 $Gyr^{-1}$ for the arms, thus avoiding an interaction (of a large magnitude) . In their fig.5, Bissantz et al showed a short theoretical arm coming at the end of the galactic bar, seen almost perpendicularly at the sun, and thus not detectable through the tangential method used here.

It is not clear if the transition from the bar to the spiral arms is of the 's' or the 'r' type, as predicted by some theoretical models . The 's' transition assumption would produce a significant pitch angle change in the inner galaxy, while the 'r' transition would not. Several numerical model assumptions may or may not be valid at some level, requiring further studies. Our observational data appear to show a favor for the 'r' type. Our Figure 6 does show a potential ring structure, linked to the spiral arm origins.

In the original model of Lin and Shu (1964), the pitch angle is determined by the dispersion relation, which would produce a change in pitch angle. Many newer density wave models have been made since, each differing in their assumptions (Dobbs & Baba 2014). Our observational data appear to favor later density-wave models, rather than the exact original model of 1964.

Modelling wise, here we use the equation for a perfect logarithmic arm, with a constant pitch angle. The assumption of perfection and constancy are employed for this galaxy as a rough simplification, as we do not have enough observations to specify parametrically all observed small deviations from this equation. A roughly logarithmic shape is well observed in many other spiral galaxies. Of course, the logarithmic shape in the Milky Way may not be strictly perfect. The constancy in pitch angle is well supported in several spiral galaxies - see Honig and Reid (2015 – sect. 5.1 and fig. 9).

In previous papers, we got the accurate galactic longitudes of the arm tangents (Paper VI; Vallée 2014c), and the accurate global arm pitch angle (Paper VIII). In our new fit, we employ these 2 accurate results in our modelling. The arms are assumed to follow a logarithmic spiral to the arm tangents in galactic longitudes between 4 and 8 kpc from the GC, one gets a good fit by varying the parameters for the start of the spiral shape ($r_o$ and $\theta_o$) in the x-y disk plane (belonging to the same single spiral). As used elsewhere (Vallée 1995 – his equ. 1; Vallée 2002 – his equ. 1; Vallée 2015 – his fig.1), we employ a (x,y) coordinate system with the origin at the Galactic Center (where x=0 and y=0), the y-axis toward the Sun (0, 8.0), the x-axis to the right (perpendicular to the y-axis, parallel to galactic longitude 90 degrees), then $\theta$ is the angle measured counterclockwise from the x-axis (thus the Sun has $\theta$ of 90 degrees from the x-axis). Here $\theta_o$ is the angle where a spiral arm starts.

We find the best fit for a fixed $r_o$ by varying $\theta_o$, then taking another fixed $r_o$ and varying $\theta_o$, on a grid of all $r_o$ and $\theta_o$. When linking these best fits on a ($r_o$, $\theta_o$) plane, a line appears. Thus we found a best fit anywhere along the line ($r_o$ ; $\theta_o$) = (2.0 kpc ; -40°) to (2.4 kpc ; 0°) in the $r_o$ - $\theta_o$ plane. No other parameters in that gridded plane could get a higher fit. The uncertainties in the parameters were ±0.02 kpc for $r_o$, and ±2 degrees for $\theta_o$.

**Figure 2** shows the best fit for an arm pitch angle of -13.1°, with the parameters ($r_o$ ; $\theta_o$) = (2.2 kpc ; -30°). We used as the standard a distance of 8.0 kpc for the Sun to GC. We adhered to a flat rotation curve, with the orbital velocity of 220 km/s, for a distance estimate (see Vallée 2008).

From this figure, the new fit predicts the location of two new arm tangents: the arm location in the inner Galaxy, notably the arm tangents at galactic longitude +20° ± 0.5° for the 'start of the Norma arm', and at -17° (=343°) ± 0.5° for the 'start of the Sagittarius arm'.

We tested our model by changing the arm pitch angle to p= -12.8° and fitting the other arms does not change this predicted Sagittarius arm at -17° longitude, to within ± 0.5°.

Many complex theoretical models exits. More complex models may fit the limited observational data to a good extent, yet having more parameters is not a gauge of physical reality (Occam's Razor requires the simplest model that can fit the available observations), as their predictions outside the limited observational range may deviate from physical reality. Thus the complex numerical SPH models of Bissantz et al (2003) are fascinating, and some may be right, but many parts are assumed (not observed). They assumed a variable circular rotation velocity curve (their Fig.1 shows a kind of parabola, thus 3 parameters are used) instead of one flat value as used in our model here. They assume an arbitrary start of the galactic arms at 3.5 kpc (their section 2.2), instead of our 2.2 kpc value here. They assumed a main thick bar length of 3.5 kpc (end of their sect.2.1) instead of our 4.2 kpc value here in our Fig.4, etc. Bissantz et al used a dark matter halo to offset a velocity deficit in their standard model (their Section 3). For our simpler model it is not necessary at this stage to fit dark matter to the observations listed in our manuscript.

The observational paper by Green et al (2011) attempted to delineate the inner Galactic structure. They emphasized the 6.7 GHz methanol maser data (indicative of young stars) over the data from many other arm tracers (dust, cold gas, etc) that we use here (see our Tables 4 and

5). An arm definition ought to include all these different tracers, to differentiate it from a density enhancement from an interarm cloud having some young stars and their masers. Their maser data are smoothed on small scales and binned on the larges scales to identify the denser maser regions, assuming an interpretation of denser maser regions as the starts of spiral arms, the galactic bar, and the so-called 3kpc features. Their complex model (their fig. 5 and table 2) is fascinating, and many parts are assumed. They assigned distance based on a flat rotation curve of 246 km/s, instead of the value of 220 km/s used in our model. They choose the end of the long, thin galactic bar (their fig.5) at a half-radius of 3.4 kpc (their sections 3.1.2 and 3.2.4). They assumed a short, thick bar half-length of 1.9 kpc (their fig.5). Our values are different, 4.2 kpc and 2.1 kpc here in Table 3 for the long and short half-lengths. They used the Sun to galactic center distance as 8.4 kpc, instead of the 8.0 kpc value here. They assumed an arbitrary start of the galactic spiral arms at 4.0 kpc (their section 3.2.4), instead of our 2.2 kpc value here. Their arm modelling (their Fig.1) stays within longitude $l = \pm 28^o$ of the Galactic Center, and they did not attempt to extrapolate their arms outside and around the sun $(28^o < l < 232^o)$. Our model here attempts to fit <u>together</u> both the inner longitudes as well as outer longitudes where tangents to the arms are observed $(280^o < l < 60^o)$.

     Green et al (2011) fitted two arms to the ends of the long, thin galactic bar (Scutum arm, Perseus arm – their sections 3.3.1 and 4). There is a problem for the model of Green et al which fit arms at the ends of the long thin bar, because observations showed some arm tangents at galactic longitudes very close to the Galactic Center (at a galactocentric radius less than that at the end of the long bar). Thus the arm origin for Sagittarius, as obtained from its galactic tangent at -17 degrees (343 degrees), is at a galactocentric radius of 2.4 kpc, while the end of the long, thin bar is now at 4.2 kpc (table 3 here). This arm cannot both start at the end of the long bar and proceeds inward toward the Galactic Center and outwards towards the Sun. Similarly, the galactic tangent of the Norma arm at +20 degrees is at 2.8 kpc, smaller than that of the ends of the long thin bar at 4.2 kpc. Hence their model is physically impossible, if the long thin bar is the start of the physical arm. This arm cannot both start at the end of the long bar and proceeds inward toward the Galactic Center. My model here is to fit the thick short bar ending at 2.1 kpc to the start of two spiral arms, seen in tangents at a radius greater than that value (2.4 kpc): Perseus arm, Scutum arm (fig.4). It follows by deduction that the long, thin bar may be a composite, much like the so-called blurred molecular ring at a galactic radius of 3.7 kpc (section 4.3). Some models predict that the long thin bar may be a composite of other segments (leading ends, s-trails), thus not a single real physical thing (section 4.2). The long thin bar, as well as the blurred molecular ring, could not be a real physical barrier, and not able to stop the spiral arms.

     Earlier, Vallée (2008) used a simple model (his fig. 1 and table 2) to fit arm tangents close to the Sun. Some parts were assumed. Thus the distance from the Sun to the Galactic Center (7.6 kpc) and a short, thick bar half-length (3.1 kpc) and a start of the spiral arms (3.1 kpc) with an arm pitch (-12.8 degrees) were employed in a model to fit the observed arm tangents located close to the Sun (his table 2). In the updated model here, we use the distance from the Sun to the Galactic Center (8.0 kpc) and a short, thick bar half-length (2.1 kpc) and a

start of the spiral arms (2.1 kpc) with an arm pitch (-13.1 degrees). This new arm pitch, from the more comprehensive data of Table 7 in Vallée (2015), and the better defined bar half-length (from the more comprehensive data in Table 3 here), is now employed here in order to extrapolate better the spiral arms on the the way down towards to the Galactic Center.

## 4. Compilation and statistics

Here we seek to assemble the main observational clues **(bulge, bars, ring) that could help us later (Section 5) to tie our arm fit (Section 3) to the inner Galaxy.**

4.1 The central bulge of red clump stars and the 'nuclear' bar

The size and orientation of the nuclear bulge around the Galactic Center are difficult to measure. It can be defined as a spheroid within galactic longitudes $|l| < 20^o$ and galactic latitudes $|b| < 10^o$ (Dwek et al 1995). It is not an axisymmetric spheroid, but displays a bar-shape with its near end in the first Galactic quadrant (the barred bulge model).

Using optical photometry, red clump (RC) giant stars bordering the edge of the galactic bulge were observed aby Nataf et al (2015), who inferred a full bulge size of roughly 2.1 kpc vertically across the galactic plane (their Fig.8a), about 1.1 kpc horizontally along the galactic longitudes (at a mean distance of 8.4 kpc), and about 2.4 kpc in depth along the line of sight near longitude $l = 0^o$ (their Fig. 8b), with an orientation of the long axis to be at $30^o$ from the line of sight to the Galactic Center (their Section 4.1).

Similarly, Qin et al (2015) used red clumps and inferred a full bulge size of roughly 2.1 kpc across the Galactic plane (their Fig. 15a), about 1.4 kpc along the galactic longitudes (at a mean distance of 8.5 kpc; their Fig.1d), and about 2.2 kpc along the line of sight near longitude $0^o$ (their Fig. 15a), with an orientation of the long axis of about $20^o$ from the line of sight to the Galactic Center (their Sect.2).

For a sun-to-GC distance of 8.0 kpc, their averaged parameters for Red Clump stars can be fitted roughly by a narrow box at the Galactic Center, inclined at $25^o$ to the line of sight to the Earth, with a full length of 2.5 kpc long (giving a line of sight of 2.3 kpc at an angle of $25^o$ and a width of 1.1 kpc along the galactic plane), and with a height of 2.1 kpc across the galactic plane. Externally linking these ends in the disk plane would form a diamond-shaped bulge.

Numerous studies have been published about a small 'nuclear bar', within $4^o$ of the Galactic Center (within 0.6 kpc), in Galactic quadrant IV (orientation greater than $90^O$) – see Fig.3 in Alard (2001); Fig.4 in Nishiyama et al (2005); Fig.8 in Sawada et al (2004). The nuclear bar shows a radial half-length near 0.2 kpc and orientation of $100^o$ to the line of sight, thus roughly along the minor axis of the bulge (Sect. 4.2 and Table 2 in Rodriguez-Fernandez & Combes 2008), or of half- length 0.5 kpc oriented at $70^o$ (Sect. 3.1.2 in Sawada et al 2004), or of half-length 0.5 kpc (Sect. 2.2 in Lopez-Corredoira et al 2001a). Other models did not need a nuclear bar (Gerhard & Martinez-Valpuesta 2012).

**Table 2** displays these results, and more. One could take a median average of these parameters, yielding a radius of 0.5 kpc at $100^o$ to the line of sight.

4.2 The 'galactic bar' and the 'long bar'

Numerous studies have been made of the 'galactic bar', the main axis of the galactic bulge, also termed the barred bulge or the 'short' bar. Its orientation appears to be between $20^o$ and $30^o$ from the line of sight from the sun to the Galactic Center.

In addition, beyond the galactic bar, an overdensity of stars in the galactic disk plane was found later, with a rather narrow lateral extent along the line of sight – it was termed the 'long bar' (e.g., Gerhard & Wegg 2014). Its orientation appears to be between $30^o$ and $40^o$ from the line of sight from the sun to the Galactic Center (longitude $l=0^o$).

**Table 3** catalogs some recent published results on this topic. One could easily adopt for the short bar a half-length of 2.1 kpc and position angle of $25^o$, and for the long bar a half-length of 4.2 kpc and position angle of $35^o$.

Some groups pointed out that the misalignment of two central bars would be difficult to understand dynamically, and could be transient; two nearby independent and misaligned bars with sizes differing by a small factor would experience a mutual gravitational attraction (e.g., Gerhard & Wegg 2014). A theoretical group has claimed a single, thick, short bulge bar, plus a second, thin, long bar, with all axes having the same orientation to the line-of-sight to the Sun (Fig. 1 in Athanassoula 2012; Fig.1 in Romero-Gomez et al 2011).

Other groups have proposed a model with a bar and a ring or an S-trail. A central bar, *rotating* in galactic longitude, could develop a segment at its 2 edges, to give an outer ring (Fig.2 in Romero-Gomez et al 2011), or to give an overall S-shaped trail (Fig.17 in Wegg et al 2015). A model with 2 bars could develop a physical link between the outer edges of the short bulge bar and those of the long bar, appearing as 'leading ends' (Figure 1 in Martinez-Valpuesta & Gerhard 2011); Fig.1 in Martinez-Valpuesta 2013).

For comparison, other nearby spiral galaxies show a long bar misaligned with the bulge major axis: NGC 2217, NGC 3992, NGC 4593 (Compère et al 2014).

4.3 The blurred 'molecular ring'

Earlier observations with radio telescopes having a low angular resolution, looking near the Galactic Center, had found a fair amount of molecules there, between 4 and 5 kpc from the Galactic Center (using a sun-to-GC distance of 8.5 kpc)

Later observations with radio telescopes having a higher angular resolution found the molecules in specific areas, but not in a whole ring. The peaks of the molecules were bunched in or close to the starts of 4 spiral arms, without much in between (Vallée 2008, his section 3.2; Dobbs & Burkert 2012, their fig.1c). Once one removes the origins of each of four arms in that radial range, then very little gas is left there in velocity space (Dobbs & Burkert 2012; Rodriguez-Fernandez & Combes 2008). Converting the radial extent of the ring for the sun-GC distance used here of 8.0 kpc, one gets the blurred molecular ring located between 3.8 and 4.7 kpc from the GC.

**Figure 3** shows a sketch of the central galactic area, within a radius of 5 kpc, with the three bars, the bulge, the blurred molecular ring.

## 5. Comparing the model arm origins with the observed objects

Here we seek to tie the 4-arm model (fig.2) to these large observed objects (fig.3).

**Figure 4** shows an overlay between the model arms (fig. 2) and these large observed objects. In the inner Galaxy, we show the so-called blurred molecular ring in black (at radial points between 3.8 and 4.7 kpc), the three bars in gray (nuclear, bulge, long), and a rough sketch of the bulge (represented broadly a red diamond).

The observed 'start of Perseus' arm, or Perseus arm origin, is very well documented, between galactic longitudes -23° (= 337°) for the $^{12}$CO 1-0 peak intensity and -20° (=340°) for the hot dust lane (Table 1 in Vallée 2014c). This arm is just outside the zone being investigated here.

Similarly, the observed Scutum arm is well documented, between galactic longitudes +26° (dust lane and methanol masers) and +33° for the $^{12}$CO 1-0 peak intensity (Table 1 in Vallée 2014c). This arm is also just outside the zone being investigated here.

Comparing with the observed data, we also find these.

(i) The fit predicts the location of the origins of these two spiral arms near the minor axis of the bulge: the start of the Norma arm (in the first Galactic Quadrant) and the start of the Sagittarius arm (in the 4$^{th}$ Galactic Quadrant).

(ii) The fit predicts the location of the origins of the other two spiral arms near the major axis of the bulge: the start of the Perseus arm (in the first Galactic Quadrant), and the start of the Scutum arm (in the 4$^{th}$ Galactic Quadrant).

(iii) The fit shows that the 'long bar' crosses the start of the Norma arm and the start of the Sagittarius arm, without apparently interacting with them, hence we deduct that it may not be a physical barrier. The long thin bar may be a composite of many segments, much like the so-called blurred molecular ring at a galactic radius of 3.7 kpc (section 4.3). In the literature we find already some models where the long thin bar may be a composite of many segments (leading ends, s-trails), thus not a single physical thing (section 4.2). The long thin bar, as well as the blurred molecular ring, could well not be a real physical barrier, and would be incapable to stop the spiral arms.

One could regroup the many published theories, as follows.

5.1 Same pitch angle, tying to bars

A model with an equidistant arm origin exists. Englmaier & Gerhard (1999, their sect. 4.1) modeled the gas dynamics in the galactic disc, and looked for quasi-equilibrium flow solutions. In their Fig.7, outside the immediate bar area, one arm emanates near one side of the *major* axis of the galactic bar and another arm from near the other side, while two arms originate from near the *minor* axis of the bar (one on each side). Ditto in Fig.5 in Bissantz et al (2003). The models by Bissantz et al do not use perfect logarithmic arms, so it is not clear to what extent their gas model apply to the Milky Way.

Models by Bissantz et al (2003 – Sect. 6) have a bar pattern speed of 60 /Gyr and a spiral arm pattern speed of 20 /Gyr, so the 4-arms exhibit a back and forth oscillation in the bar frame, and are connected to the interior zone via a transition region with a radius near 3 kpc.

In our Fig. 4 here, we can see that the 'start of the Sagittarius' arm and that the 'start of the Norma' arm are along the extrapolation of the bulge's *minor* axis; likewise, the 'start of the Perseus' arm and the 'start of the Scutum' arm seem to be near the bulge's *major* axis.

In our Fig.4 here, we could have extended the arms originating near the minor axis of the bar (Sagittarius and Norma), until they stopped at the major axis of the bulge bar. If we did, the Sagittarius origin would fall about 1 kpc below the bar end where Perseus starts, and the Norma origin would fall about 1 kpc below the bar end where Scutum starts. We did not find theoretical models in the literature to support 4 arms out of a bar, 2 from the ends and 2 from 1 kpc below the ends.

A variant of that equidistant arm model exists. Roman-Duval et al (2009) used a 4-arm model, arbitrarily choosing to draw two arms coming from near each bar end (their Fig.11): the adjacent Scutum and Norma arms from near the bar end, and the adjacent Sagittarius and Perseus arms from near the other bar end. The two arms coming from near the same bar end are separated by about 1.4 kpc (for $r_{sun}$= 8.0 kpc), both having the same arm pitch angle of -12.7$^o$.

5.2 Different pitch angles.
In our Fig. 2 here, we did not try different pitch angle values for each arm, as observations of nearby disk galaxies suggested a roughly constant mean pitch (fig. 1, and Sect. 3.1).

Some arm models arbitrarily choose to draw the origins of 2 arms from each end of the galactic bar, each arm having its own different initial pitch angle. Thus in Churchwell et al (2009, their fig.16), the far end of the bulge bar is joined by two arms and an elongation (the Perseus spiral arm, the Sagittarius spiral arm, and the 'far 3kpc arm' elliptical feature), while the near end of the bulge bar does the same (the Scutum spiral arm, the Norma spiral arm, and the 'near 3kpc arm' elliptical feature). In these more complex models, each of these 4 arms and 2 elongations has its own pitch angle, but some pitch angle are not held fixed but are varied farther out, in order to match observed spiral arm tangencies.

## 6. Arm tracers in the inner galactic longitudes   $-20^o < l < +23^o$

Here we seek additional observational clues (arm tracers), in order to see if our predicted arm origins (fig. 2) can be tied.

Here we seek the imprint of the spiral arms in the galactic longitudes between -20$^o$ and +23$^o$ of longitudes. Attempts to determine the inner Galactic structure have often proven to be controversial. What is new is that we have a very well fitted arm model, using an accurate arm pitch angle and accurate arm tangents, that we extrapolate inward, telling us where to look for the origins of each spiral arm (fig. 4).

A false spiral arm could be due to a misinterpretation of an intensity peak as one goes along the galactic longitude. This occurs when either (i) the peak intensity is due to an emission from a small sphere (globular clusters, supernovae, etc) but with little emission along the line of sight (tube), or when (ii) the peak intensity comes from an integration over a large radial velocity band encompassing more than one object at vastly different distances along the line of sight (cutting 3 arms, say). Possible examples of a false spiral arm are listed in **Appendix B**, when appropriate.

From Figure 4, our computer model predicts that the 'start of the Sagittarius' arm (with CO at the mid arm) would be at l= -17$^o$ (=343$^o$), and that the 'start of the Norma' arm (with CO at the mid arm) would be at longitude l= + 20$^o$. Other tracers should be offset from the mid-arm (CO), appearing at shorter longitudes in Galactic quadrant I, and at longer longitudes in Galactic quadrant IV.

In **Table 4**, we catalog observed peak intensity from various arm tracers, around these inner longitudes, from the literature.

6.1 At negative longitudes.
The observed 'start of the Sagittarius arm' at negative longitudes can be seen from l= -8$^o$ to l= -17$^o$. From figure 2, the computer model for the origin using CO for the middle of the Sagittarius arm is at l= -17$^o$. This would cut over the far end of the 'long bar'.

In Green at al (2011), the "Carina-Sgr origin" arm is indicated at l= -8$^o$ (their Fig.3), as identified by a very high density of masers. Green et al (2011 – their table 1) also listed a moderate methanol maser concentrations between l= -6$^o$ to l= -11$^o$, which they identified as associated with the Sagittarius arm. If masers are located at the inner edge of the arm, then the middle of the Sagittarius origin arm (in broad $^{12}$CO) could be at l= -17$^o$, for the usual offsets seen elsewhere and as pointed out in Vallée (2014a; 2014c).

*Thus it seems that the location of the so-called 'start of the Sagittarius' arm seen in masers near l= -6$^o$ to -11$^o$ is in fact the 'inner edge' of the 'start of the Sagittarius' arm, whose mid-arm is computed near l= -17$^o$.* The interarm distance at the Sun, between the Sagittarius and the Perseus arm is about 3 kpc at the Sun. The interarm distance diminishes somewhat as we go down towards the Galactic Center. Also in that interarm near the Sun, there is the Orion 'spur', or 'armlet', or small 'ridge', and we surmise that some similar spurs may exist elsewhere in our galactic disk. The true inner edge of that arm may be at l= -11 degrees, some 6 degrees offset from the midarm, hence giving a linear offset of about 800 pc. Theoretically, one would not expect the arm to start abruptly so near to the minor axis of a bar. The longitude range indicated (5 degrees) may suggests 'roots', narrowing and converging to reach the start of the arm (as in an Y shape, roots at the top, arm at the bottom). In addition, this longitude range as indicated (5 degrees) may harbor some 'spurs', like the one near the Sun.

6.2 At positive longitudes.

The start of the Norma arm at positive longitudes can be seen from l= +12$^o$ to l = +20$^o$. From this figure 2, the predicted origin of middle of the Norma arm is at l= +20$^o$. This would cut over the near end of the 'long bar'. Changing the arm pitch angle to p= -12.8$^o$ and fitting the other arms does not change this predicted Norma arm at +20$^o$ longitude, to within 0.5$^o$.

In Green et al (2011), the 'Norma origin' arm is indicated at l= +12$^o$ (their Fig.3); this location is identified by a high density of masers (their Sect.4). Also, Green et al (2011) found a moderate density increase between l= +10$^o$ to +14$^o$, which they attributed to a galactic bar (with a question mark), as well as between l= +18$^o$ to l= +22$^o$. However, we note that it does fit with our CO model at l= +20$^o$ (fig. 2).

In our computer model, the middle of the 'start of the Norma' arm could be at l= +20$^o$ with the masers located at the inner edge of the arm, some degrees away from the mid arm observed with broad $^{12}$CO as pointed out in Vallée (2014a; 2014c).

*Thus it seems that the location of the so-called 'start of the Norma' arm seen in masers near l= +10$^o$ to +14$^o$ is in fact the 'inner edge' of the 'start of the Norma' arm whose mid-arm is near l=+20$^o$.* The true inner edge of that arm may be at l= +14 degrees, some 6 degrees offset from the midarm, hence giving a linear offset of about 800 pc. Theoretically, one would not expect the arm to start abruptly so near to the minor axis of a bar. The longitude range indicated (4 degrees) may suggests 'roots', narrowing and converging to reach the start of the arm (as in an inverted Y shape, roots at the bottom, arm at the top). In addition, this longitude range as indicated (4 degrees) may harbor some 'spurs', like the one near the Sun.

We thus have a confirmation (via masers) of the origins of two spiral arms (computed for the mid-arm as seen with broad CO), as per our arm definition above (narrow tube, different tracers, ordered offsets).

## 7. Compilation and statistics on the so-called '3-kpc-arm' features

Here we seek to investigate observed clues to the '3-kpc-arm' features, in order to investigate a possible link later (Section 8).

The feature known as the '3-kpc arm' has not been proven to be a true arm. It could be either a small arc, a spur, an armlet, an oval, an ellipse or a ring, or the beginning of a long arm turning behind the galactic center (e.g., Fig. 5 in Green et al 2011).

The '3-kpc-arm' feature cannot be a single spiral arm, since its tangents are almost symmetric with respect to the Sun-to-Galactic-Center line, giving it a near $0^o$ pitch angle (as would have a ring); for example, the Carina tangent near l= -76$^o$ for old HII complexes and its counterpart the Sagittarius tangent near l=+51$^o$ are combined to give a joint pitch angle near -13$^o$ (Table 1 and Equation 10 in Vallée 2015).

**Table 5** assembles various published locations for the tangent to this '3-kpc-arm' feature. The disparity in values shows that it is difficult to get an accurate estimate of a tangent to this feature.

7.1 At negative longitudes.

There is a well-known arm identified as the 'Start of the Perseus arm' at galactic longitude l= 337$^o$ (Table 2 in Vallée 2008; Table 3 in Vallée 2014a; Table 3 in Vallée 2014b; Table 1 in Vallée 2014c).

Some misidentifications of the '3-kpc-arm' feature were actually overlapping in galactic longitudes with the 'start of Perseus' arm between l=337$^o$ and l= 340$^o$. Thus Hou & Han (2015) refer to the l= -23$^o$ arm ( l=337$^o$ ) as the 'Near 3-kpc South arm' (in their Table 1).

The tangent to the "3-kpc-arm" feature is predicted to be near l= -23$^o$ (Dame & Thaddeus 2008, extrapolating from their Fig.1), to near l= -12$^o$ (Englmaier & Gerhard 1999, thick dashed curve in their fig.11). In Dame & Thaddeus (2008), its radial velocity at l=0$^o$ is near -53 km/s (their Table 1). From their figure, its radial velocity at l= -17$^o$ can be extrapolated here to about -120 km/s.

Also, Green et al (2011 – their fig.3) positioned the '3-kpc-arm' feature tangent near l= -22$^o$.

Englmaier & Gerhard (1999) positioned a tangent to the '3-kpc-arm' feature at l= -21$^o$ (their table 1) and another one at -12$^o$. In Churchwell et al (2009), using the 4.5μm star counts, the '3-kpc-arm' feature is positioned at l= -21$^o$ (their fig. 14). In Sanna et al (2014), the "3-kpc-arm" feature is drawn as an azure ellipse, being seen tangentially at l= -17$^o$ (their Fig.6). In Beuther et al (2012), the tangent to the "3-kpc-arm" feature is at l= -17$^o$ (their Fig. 3). In Green et al (2011; 2012), the "3kpc arm" is positioned using CO as between l= -14$^o$ and l=+14$^o$ (their fig.1), although no tangency is shown. Other data are listed in Table 5.

Excluding the longitude range of the 'start of the Perseus arm' (l ≤ -22$^o$ ), statistics can be done on the remaining longitude data range from -21$^o$ to -12$^o$. Table 5 shows that a mathematical average in negative longitudes, for the observed tangent to the so-called '3-kpc arm' feature, yields l= -18$^o$ ± 2$^o$ (s.d.m.).

7.2 At positive longitudes.

There is a well-known arm identified as the Scutum arm, between l=+26$^o$ and l= +33$^o$ (see Table 1 in Vallée 2014c). Some misidentifications of the '3-kpc-arm' feature were actually overlapping in galactic longitudes with the Scutum arm. Thus Hou & Han (2015) refer to the l= +27$^o$ stellar arm as the 'Near 3-kpc North arm' (in their Table 1).

One tangent to the "3-kpc-arm' feature is tentatively observed to be near l= +12°
(Englmaier & Gerhard 1999, thin dashed curve in their fig.11a), and another one near l= +24°.
From fig.1 in Dame and Thaddeus (2008), its radial velocity at l= +20° can be extrapolated here
to about +40 km/s.

Green et al (2011 – their Sect. 3.2.2) find 3 density peaks of 6.7 GHz masers (equivalent
to tangents) near l= +12°, +20°, and +26°, and the last one they attributed to the Scutum arm.
This leaves the +20° and the +12° for a possible tangent to the '3-kpc-arm' feature.

In Jones et al (2013, their Fig.7), the tangent to the '3-kpc-arm' feature is at l= +20°. In
Dame & Thaddeus (2008), the tangent to the "3-kpc-arm" feature appears near l= +23° from
$^{12}$CO 1-0 data (their Fig.1). In Sanna et al (2014), the "3-kpc-arm" feature is drawn as an ellipse,
being seen tangentially at l= 23° (their fig. 6). Englmaier & Gerhard (1999 - their Sect. 2.4) noted
that "the Scutum tangent is double in a number of tracers. The inner Scutum tangent at l= 24° is
sometimes referred to as the northern 3-kpc arm". In Beuther et al (2012 – their fig. 3), at
positive longitudes the 3kpc arm (called there the "far 3-kpc arm") runs from l=0° (merging with
the Sagittarius arm) to l= +25° (merging with the Scutum arm), but there is no tangent
anywhere. Other data are listed in Table 5.

Excluding the longitude range of the 'Scutum arm' (l ≥25°), statistics can be done on the
remaining longitude data range from l= +12° to +24°;. Table 5 shows that a mathematical
average in positive longitudes, for the tangent of the '3-kpc arm' feature, yields l= +21° ± 2°
(s.d.m.).

**Figure 5** shows the location of the mean longitudes of the 3-kpc-arm features.

## 8. On tying the spiral arm origins with the 3-kpc-arm features

Here we examine the overlap between the origins of spiral arms and the tangents (fig. 5)
to the so-called '3-kpc-arm' features (table 5).

At negative longitudes, the location of the '3-kpc-arm' feature is in fact very close to the
'start of Sagittarius' arm at l= -17°, the same within the mean error. Using data in Fig. 3d of
Vallée (2008), the 'start of Sagittarius' arm 's radial velocity at l= -17° can be extrapolated here
to about -100 km/s, with a large error.

At positive longitudes, the location of the so-called '3-kpc-arm' feature is in fact very
close to the location of the 'start of Norma' arm at l=+20°, the same within the mean error.

Using the data in fig.3a of Vallée (2008), the 'start of the Norma' arm's radial velocity at
l= +20° can be extrapolated here to about +60 km/s, with a large error.

We thus have a confirmation that the two tangents to the '3-kpc-arm' feature are
essentially at the same galactic longitudes (within 1°) as the two tangents to the start of two spiral
arms (Norma and Sagittarius). In the rough assumption of a uniform distribution over 180
degrees, the chance probability here is: $(2\text{ sdm}/180)^2$ or 0.05 %.

A symmetry case can be made. We found the observed so-called '3-kpc-arm' feature to
be attached to the 'origin of Sagittarius' arm and the 'origin of Norma' arm. Thus we can
reasonably assume that the 2 other arms must be symmetric. By symmetry, we surmise that each
of the 4 spiral arm origins has an associated '3-kpc- arm' feature, two of which can be seen as
tangents from the Sun (Table 5), while the other two would be mostly across the line of sight
from the Sun (not tangent). This is also inferred from the radial velocity offsets as observed
between -12° < l < +12° (fig. 1 in Dame and Thaddeus 2008).

Is there more? The sharing of the same spatial segments by the 'start of the arms' and the '3-kpc-arms' (fig. 5) may be a dynamically unstable situation. If that is the case for the 4 arm origins by symmetry, then some effects could be predicted:   a) by symmetry, there chould be two other '3-kpc arm' features but along the line of sight ; b) by their extent, there could be a potential link between these four '3-kpc arm' features,   giving a pseudo ring.

In **Figure 6**, we show the 2 tangents to the '3-kpc-arm' feature (orange bars), along with the starts of the spiral arms, as observed at $l= -18°$ and $l=+21°$ (table 5), and those 2 inferred by symmetry along the line-of-sight (orange bars). We also traced a potential link for these 3-kpc-arm' features, as a pseudo ring (orange triangles).

The velocity difference, between the '3-kpc-arm' feature and its associated spiral arm origin, could well suggest two gas segments in a contiguous space, rather than two gas segments at very different spatial distances.    Other theoretical models do not provide an inspiration for this contiguous space with a discontiguous velocity.

## 9. Conclusion

The global arm tracers (table 1) and accurate arm pitch angle statistics (Vallée 2015) as well as accurate galactic arm tangencies (Vallée 2014c) were employed here into our computer modelling, to get predictions for the tangent to the start of each spiral arm (fig. 2). A constant pitch angle was assumed as observed elsewhere (fig. 1).

Here we propose a new view of the Milky Way's interior ($-20° < l < +23°$), by putting together several independent observational data, and some spiral arm modelling.   Thus we looked at bars (table 2 and table 3, fig. 3). A search for inner arm tangencies allowed us to see observational tracer offsets in the inner arms (table 4).

The new model computed fit predicts the location of  the arm tangents for the start of the Sagittarius arm, and for the start of the Norma arm. We found  the model tangent to the 'start of the Sagittarius' spiral arm (middle arm) at $l= -17°$ ($=343°$) $±0.5°$, while the model tangent to the 'start of the Norma' spiral arm  (middle arm) is at $l= +20° ±0.5°$.

The new model (fig. 4) fit predicts the location of the origins of two spiral arms, near the extension of the minor axis of the bulge bar for the Norma arm origin (in the first Galactic Quadrant) and for the Sagittarius arm origin (in the 4th Galactic Quadrant), and near the major axis of the bulge for the Perseus arm origin (in the first Galactic Quadrant), and the Scutum arm origin (in the 4th Galactic Quadrant).

We found (via published masers data) the origins of two spiral arms (computed for CO), as per our arm definition above (narrow tube, different tracers, ordered offsets). Thus the location of the so-called 'start of the Sagittarius' arm seen in masers near $l= -6°$ to $-11°$  is in fact the 'inner edge' of  the 'start of the Sagittarius' arm, whose mid-arm  is near $l= -17°$. The location of the so-called 'Norma origin' arm seen in masers near $l= +10°$ to $+14°$ is in fact the 'inner edge' of  the 'start of the Norma' arm whose mid-arm  is near $l=+20°$.

The observed  tangents to the '3-kpc-arm' features (table 5)  are found statistically near $l= -18° ± 2°$ and $l= +21° ±2°$ , after excluding misidentified features overlapping with well-known spiral arms. We confirmed that the two tangents to the observed '3-kpc-arm' feature are essentially at the same galactic longitudes as the two model tangents to the start of two spiral arms (Norma and Sagittarius). The chance probability here is $(4/180)^2$ or 0.05 %.  The negative longitude location of the observed '3-kpc-arm' feature is in fact very close to the model-computed 'start of Sagittarius' arm at $l= -17°$, the same within the mean error. The positive longitude location of the observed '3-kpc-arm' feature is in fact very close to the model-computed location of the 'start of Norma' arm at $l=+20°$, the same within the mean error (fig. 5).

Comparing with some theoretical models, we surmise that the co-spatiality of the tangents of the '3-kpc-arm' features with the origins of the spiral arms may allow a dynamical interaction with one of the bars, scooping up gas from the nearby spiral arms into a pseudo ring (fig. 6).

### Acknowledgements.
The figure production made use of the PGPLOT software at the NRC Canada in Victoria. I thank E. Berkhuijsen for the fine distinction on the writing of Patrickeyev (in Poland) versus Patrikeev (in Paris). I thank the referee for a very helpful and thorough report.

### Appendix A
Comments of some global spiral arm fits. Griv & Jiang (2015) employed 242 open star clusters near the Sun, with a mean age of 550 My, and fitted different models with m=0,1,2,3,4,5 arms (their Table 1 and Fig. 8a). Their fit is over a small area near the Sun (within 3 kpc), and did not use tracers over the whole Milky Way (a radius of 16 kpc is given in their fig. 8a). . Their fig.8 does show the observed clusters to appear randomly distributed, and do not seem to align along spiral shapes. Their table 1 with the minimization of the estimator S shows the S value to be about the same (S between 482 and 486) for all arm models m=1,2,3,4,5, although they stated that m=1 'dominates' and that all models could be 'superposed' (their Sect. 5). Their table 3 shows that the lowest S value is for a model with m=5 arms, which contradicts the m =1 model as dominant.

Mel'nik et al (2015) used 674 local Cepheids near the Sun and a model with an outer ring composed of an ascending (trailing spiral) segment and a descending (leading spiral) segment (their Figs. 1 and 2). In their Fig. 2, the positions of observed nearby Cepheids appear random, not along spirals. Two rings are used: $R_1$ at 6.0 kpc (varying between 5.8 and 6.3 kpc), and $R_2$ at 8.0 kpc (varying between 7.6 and 8.5 kpc), using their $R_{sun}$=7.5 kpc. Many of the observed Cepheids are concentrated near the Sun (within 3 kpc or so), but a galactic model view covering 12 kpc by 12 kpc is shown in their Figure 2.

Hou & Han (2015) provided a 4-arm model for the gas arms, and also a 3-arm model for the stellar arms. Their stellar arm tangents come from the GLIMPSE data (3.6 to 8.0 µm) and the 2MASS data (1.25 to 2.16 µm). These results used a 'universal extinction' law, the same for all galactic longitudes. In contrast, a large variation in dust extinction along galactic longitude, from one sightline to another, was reported in the mid-infrared (2µ to 8µm) by Gao et al (2009, their Fig. 6). More dusty arms (Carina and Norma) could make all of their stars disappear in the mid-infrared.

### Appendix B
Possible examples follow of false spiral arm claims between $l=+22°$ and $l=+26°$, including the so-called '3-kpc-arm' features, when misidentified as a well-known arm.

The peak at $l= +25°$ in Fig. 5 and Tab. 6 of Russeil (2003) contains objects at very different distances along the line of sight. The warm $^{12}CO$ in Solomon et al (1985 – their Fig. 1b and fig. 2) is not attributed to a specific arm, and shows up as a diffuse 4° band between $l= +22°$ to +26°, and a diffuse 80 km/s range between $v_{lsr}$= +40 to +120 km/s. Similarly, the CO in Cohen et al (1980 – their fig. 2) has a huge radial velocity range at l= + 24°, namely from +40 to +120 km/s, encompassing many arms and objects at different distances (when using a flat rotation curve to get the radial distance, say).

Sanders et al (1985) showed in their Fig. 5a some radio HII complexes located at l=+25°, and $v_{lsr}$ = +110 km/s, very close to their narrow Scutum arm, but slightly inside their very broad '4-kpc-expanding-arm'; these HII complexes could well belong to one or to the other arms.

The '3-kpc arm' feature in Bania (1980 – fig.7) near l=+24° was not fitted to CO clouds; the fit was taken from older HI 21cm data, which were dependent on the kinematical HI model used.

The '4-kpc arm' feature in Dame et al (1986 – fig.9) near l=+26° were not fitted to CO clouds; the fit was taken from older (1972) HI 21cm data, which were dependent on the kinematical HI model used.

A double peak in the 240μm emission, due to strong absorption, can be seen around the Scutum arm at +31° and +23° in Fig. 1 of Drimmel (2000); one of the double peak may not be an arm.

## Table 1 – Recent studies of the spiral arms in the Milky Way (early-2015)

| Pitch Angle (deg.) | No. of arms | Arm shape [a] | Inter-arm [b] (kpc) | Data used | Figures and references |
|---|---|---|---|---|---|
| -15 | 4 | log | 3.7[c] | NIR stars in Embedded clusters | Fig.12 in Camargo et al (2015) |
| -13 | 4 | log | 3.2[c] | HI and $H_2$ gas | Fig.5 in Eger et al (2015) |
| -13 | 4 | log | 3.5[c] | Classical Cepheids | Fig.4 in Dékany et al (2015) |
| -21 | 4 | irreg | 2.8 | CO maps, smooth particle HD | Fig.10c in Pettitt et al (2015) |
| -13 | 4 | log | 3.2[c] | λ1.1mm dust continuum | Fig.15 in Ellsworth-Bowers et al (2015b) |
| -13 | 4 | log | 2.8 | Photometry of open star clusters | Fig.15 in Giorgi et al (2015) |
| 0 | - | ring | 2.6[c] | Classical Cepheids | Fig.2 in Mel'nik et al (2015) |
| -14 | 4 | log | 3.2[c] | methanol masers | Fig.5 in Krishnan et al (2015) |
| -14 | 4 | log | 3.5[c] | GeV cosmic ray | Fig.2a in Kissmann et al (2015) |
| -13.5 | 4 | log | 3.2[c] | Far infrared large filaments | Fig.3 in Wang et al (2015) |
| -12 | 4 | polyn | 2.7[c] | Radio-Polarization and RM | Fig.12b in Purcell et al (2015) |
| -4.3 | 1 | log | 3.6[c] | Open star clusters | Fig.8a in Griv & Jiang (2015) |
| -10 | 4 | log | 3.0[c] | astrometric masers | Fig.1 in Sakai (2015) |
| -10 | 4 | log | 2.6[c] | Classical Cepheids | Fig.9 in Dambis et al (2015) |
| -12.5 | 4 | log | 3.0 | 870µm dust & $NH_3$ sources | Fig.18 in Wienen et al (2015) |
| -11 | 4 | log | 3.1[c] | water masers | Fig.4 in Sakai et al (2015) |
| -13 | 4 | log | 3.4[C] | λ1.1µm dust continuum | Fig.8 in Ellsworth-Bowers et al (2015a) |
| -12 | 4 | log | 3.4[c] | globular clusters | Fig.3 in Rossi et al (2015) |
| -13 | 4 | log | 3.1[c] | MIR stars in GLIMPSE/2MASS | Fig.. 9 in Hou & Han (2015) |
| **-13** | **4** | **log** | **3.2** | **Median value (all data)** | |
| -13 | 4 | log | 3.2 | Unweighted mean (excluding the rings of Mel'nik et al 2015) | |
| ±1 | - | - | ±0.2 | Standard deviation of the mean (all data) | |

Notes:
(a): The arm shape can be logarithmic (log), polynomial (polyn), ring, irregular (irreg), or complex.
(b): The separation between the Perseus arm and the Sagittarius arm, through the Sun's location.
(c): Corrected for 8.0 kpc as the Sun - Galactic Center distance (not the 8.5 value or else as used).

## Table 2 – The nuclear bar

| Pos. Ang. (o) | radius[a] (kpc) | Data used | Reference |
|---|---|---|---|
| 100 | 0.3 | old stars in H and K | Alard (2001 – fig.3) |
| -   | 0.6 | stars in J, H, and K | Nishiyama et al (2005 – Fig. 4) |
| 70  | 0.5 | CO and OH gas | Sawada et al (2004 – Sect 3.1.2) |
| 100 | 0.2 | star counts | Rodriguez-F. & Combes (2008 – Tab.2) |
| -   | 0.5 | star counts | Lopez-C. & et al (2001a – Sect. 2.2) |
| 90  | 0.4  | mean | |
| ±8  | ±0.1 | standard dev. of the mean | |
| 100 | 0.5  | median | |

Notes:
a) Scaled back to a Sun-to-Galactic center distance of 8.0 kpc

**Table 3 – A selection of observed and computed galactic bars (since 1995), excluding the 'nuclear bar'.**

| Short, boxy, galactic bar | | Long bar | | Data used | Reference |
|---|---|---|---|---|---|
| Pos. Ang. (o) | radius (kpc) | Pos. Ang. (o) | radius (kpc) | | |
| 20 | 2.3[a] | - | - | NIR stars | Tab. 4 in Dwek et al (1995) |
| - | - | 40 | 3.9 | 2μm star counts | Sect. 6.2 in López-Corredoira et al (2001b) |
| 20 | 2.0 | - | - | gas | Fig.5 in Bissantz et al (2003) |
| - | - | 43 | 3.9 | star counts | Fig.5 in López-Corredoira et al (2007) |
| 27 | 1.2 | - | - | star counts | Tab. 1 in Rodriguez-Fernandez & Combes (2008) |
| - | - | 20 | 3.8[a] | stellar kinematics | Sect.3 in Shen et al (2010) |
| - | - | 45 | 3.2[a] | 6.7 GHz masers | Sec.3.1 in Green et al (2011) |
| 25 | 2.0 | 43 | 4.5 | 1M particles | Fig. 2 in Martinez-V. & Gerhard (2011) |
| 20 | 3.3 | 40 | 4.2 | manifolds | Sec. 2.1 in Romero-Gómez et al (2011) |
| 25 | 2.2 | 43 | 4.0 | 1M particles | Fig.1 in Martinez-Valpuesta (2013) |
| 27 | 1.4[a] | - | - | red clump stars | Fig.17 & Sect.10 in Nataf et al (2013) |
| 20 | 2.8[a] | - | - | 12M particles | Fig.4 in Antoja et al (2014) |
| 25 | 3.0 | - | - | red clump/giant stars | Fig. 6 in Bobylev et al (2014) |
| 27 | 2.0 | | | density models | Sect.3 in Gerhard & Wegg (2014) |
| - | - | 78 | 5.7[a] | diffuse gamma-ray | Fig.4 in De Boer & Weber (2014) |
| 15 | - | - | - | stellar radial velocity | Sect. 3.1 in Gardner et al (2014) |
| 20 | 2.9[a] | 20 | 4.2[a] | 8M particles | Fig.12 in Romero-Gomez et al (2015) |
| - | - | 27 | 4.4[a] | 0.7M particles | Sec.2.1 in Portail et al (2015) |
| - | - | 30 | 4.4[a] | red clump giants | Sect. 6.3 in Wegg et al (2015) |
| 30 | 1.4[a] | - | - | red clump giants | Fig.8 in Nataf et al (2015) |
| 20 | 1.2[a] | - | - | red clump giants | Fig.15 in Qin et al (2015) |
| 21 | 1.6[a] | - | - | RR Lyrae stars | Fig. 7 in Pietrukowicz et al (2015) |
| 23 | 2.1 | 34 | 4.2 | | mean (excluding the data from de Boer & Weber 2014) |
| ±2 | ±0.2 | ±3.3 | ±0.1 | | standard dev. of the mean |
| 23 | 2.1 | 40 | 4.2 | | median (excluding the data from de Boer & Weber 2014) |
| 25 | 2.1 | 35 | 4.2 | | adopted |

Notes:
a) Scaled back to a Sun-to-Galactic center distance of 8.0 kpc.

# Table 4 – Offset of each spiral arm tracer from $^{12}$CO, in the inner Milky Way ($-17° < l < +20°$)

| Arm Name | Chemical tracer | Gal. longit. $l$ [a] (°) | Angular dist. [b] to mid-arm (°) | Linear offset [c] (pc) | References |
|---|---|---|---|---|---|
| 'Start of Sagittarius' arm (Sagittarius origin): | | | | | |
| | CO mid-arm | -17 | 0 | 0 pc, at 7.5 kpc | this paper – fig.1 |
| | 870 μm dust | -17 | 0 | 0 pc | Beuther et al (2012 – fig.3) |
| | $^{12}$CO at 8' | -16 | 1 | 130 | Grabelsky et al (1987 – fig.6) |
| | 6.7GHz masers | -16 | 1 | 130 | Sanna et al (2014 – fig.6) |
| | $^{26}$Al | -14 | 3 | 390 | Kretschmer et al (2013 – Fig.9a) |
| | J, H, K star counts | -12 | 5 | 650 | Benjamin (2008 – Fig.2) |
| | 6.7GHz masers | -11 | 8 | 1050 | Green et al (2011 – table 1) |
| 'Start of Norma' arm (Norma origin): | | | | | |
| | CO mid-arm | +20 | 0 | 0 pc, at 6.5 kpc | this paper – fig.1 |
| | Some 6.7GHz masers | +20 | 0 | 0 | Green et al (2011 – table 1) |
| | 4.5 μm star counts | +19 | 1 | 110 | Benjamin (2008 – Fig.2) |
| | J, H, K star counts | +19 | 1 | 110 | Benjamin (2008 – Fig.2) |
| | Radio continuum 5GHz | +16 | 4 | 450 | Hayakawa et al (1981 – fig2b) |
| | Many 6.7GHz masers | +12 | 8 | 900 | Green et al (2011 – table 1) |

Notes:
(a): Negative galactic longitudes are converted to a positive range in a circle by adding 360°.
(b): Separation inside arm, from CO model mid-arm; positive towards the arm's inner edge (toward the Galactic Center).
(c): Linear offset, after converting the angular separation at the arm segment's distance from the Sun (taken from Fig. 2 here).

## Table 5 - Observed tangents to the so-called '3-kpc-arm' features[a]

| Feature Name[a] | Chemical tracer | Gal. longit.[b] | References | Note [c] [d] |
|---|---|---|---|---|
| So-called '3-kpc-arm' feature at negative longitudes, misidentified, part of another well-known arm: | | | | |
| | CO 1-0 | -24 | Bronfman et al (1989 – fig. 9) | In 'start of Perseus' arm |
| | CO 1-0 | -23 | Dame & Thaddeus (2008 – fig.1) | In 'start of Perseus' arm |
| | CO 1-0 | -23 | Alvarez et al 1990 – tab.4) | In 'start of Perseus' arm |
| | CO 1-0 | -22 | Garcia et al (2014 – fig. 11 & tab.3) | In 'start of Perseus' arm |
| | Masers | -22 | Caswell et al (2011 – Sect. 4.6.1) | In 'start of Perseus' arm |
| | Masers | -22 | Green et al (2011 – fig.3) | In 'start of Perseus' arm |
| | Stars (1-8 µm) | -22 | Hou & Han (2015 – table 1) | In 'start of Perseus' arm |
| '3-kpc-arm' feature at negative longitudes, outside of the 'start of Perseus' arm area: | | | | |
| | 4.5µm stars | -21 | Churchwell et al (2009 – fig. 14) | |
| | NIR | -21 | Englmaier & Gerhard (1999 – their tab.1) | |
| | Dust 240µm | -21 | Drimmel (2000 – fig.1) | |
| | 408 MHz synch | -20 | Beuermann etal (1985 – fig. 7a) | |
| | HI | -19 | Jones et al (2013 – fig.7) | labelled 'near' |
| | Massive stars | -19 | Reid (2012 – fig.3) | |
| | Masers | -16 | Sanna et al (2014 – fig.6) | |
| | 870 µm dust | -17 | Beuther et al (2012 – fig.3) | |
| | 6.7 GHz masers | -11 | Green et al (2011 – tab.1) | labelled 'far' |
| | CO 1-0 | -12 | Englmaier & Gerhard (1999 – their fig. 11a) | |
| | | ----- | | |
| | | -17.7 | mean | |
| | | ±2 | standard dev. of the mean | |
| '3-kpc-arm' feature at positive longitudes, outside of the Scutum arm area: | | | | |
| | $^{13}$CO 1-0 | +24.4 | Hou & Han (2015 – table 1) | |
| | $^{12}$CO 1-0 | +24 | Englmaier & Gerhard (1999 – sect. 2.4); **Hou & Han (2015 – table 1)** | |
| | 870 µm 19" cores | +23.8 | Hou & Han (2015 – table 1) | |
| | NH$_3$ 1-1 2' cores | +23.7 | Hou & Han (2015 – table 1) | |
| | $^{12}$CO 1-0 | +23 | Dame & Thaddeus (2008 – fig.1) | |
| | HII regions | +23 | Sanders et al (1985 – fig.5a); Hou & Han (2015 – table 1) | |
| | Masers | +23 | Sanna et al (2014 – fig.6) | |
| | Masers | +20 | Green et al (2011 – sec. 3.2.2) | |
| | $^{12}$CO 1-0 | +12 | Englmaier & Gerhard (1999 – their fig. 11a) | |
| | Masers | +12 | Green et al (2011 – sec. 3.2.2) | |
| | | ----- | | |
| | | +20.8 | mean (excluding extrapolated data) | |
| | | ±2 | standard dev. of the mean | |
| So-called '3-kpc-arm' feature at positive longitudes, misidentified, part of another well-known arm: | | | | |
| | Stars (1-8 µm) | +27 | Hou & Han (2015 – table 1) | In Scutum arm; or two arms |
| | $^{12}$CO 1-0 | +26 | Cohen et al (1980 – fig.3) | In Scutum arm; or two arms |
| | HII complex | +25 | Russeil (2003 – Tab.6) | In Scutum arm; or two arms |
| | HII complex | +25 | Sanders et al (1985, fig.5a) | In Scutum arm; or two arms |

Notes:
(a): Sometimes called also the '4-kpc arm' feature.
(b): Negative galactic longitudes are converted to a positive range in a circle by adding 360°.
(c): The 'start of Perseus' arm is well documented between galactic longitudes 337° < l < 340° in Table 1 of Vallée (2014c).
(d): The Scutum arm is well documented between galactic longitudes 26° < l < 33° in Table 1 of Vallée (2014c).

**Figure captions**

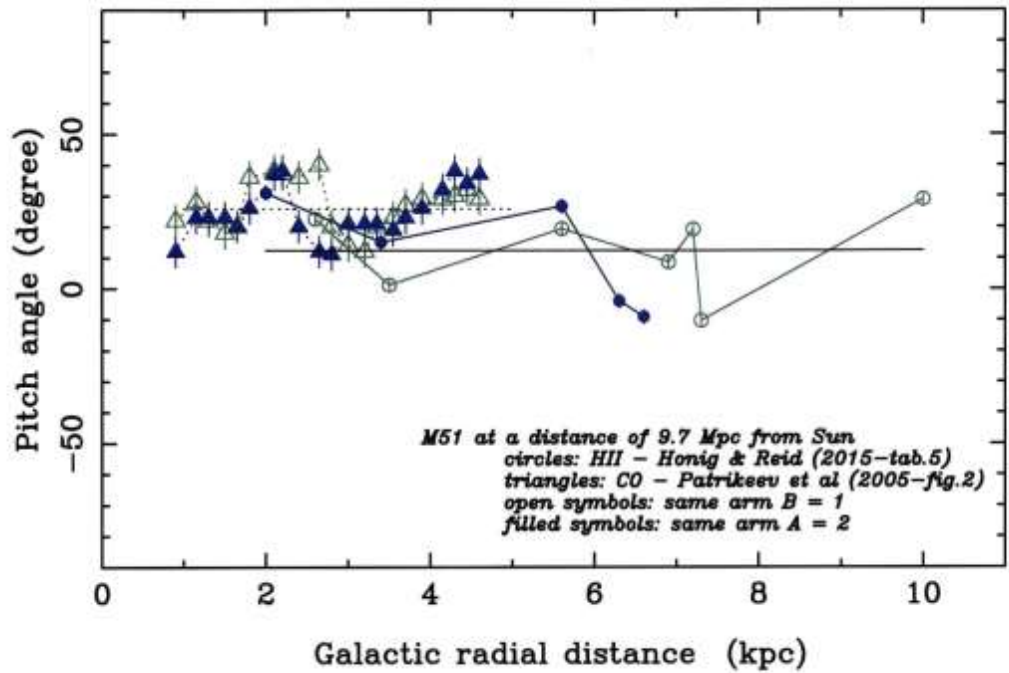

Figure 1. Pitch angle data versus galactic radii, from the literature on the spiral galaxy M51. Different authors made different assumptions about the distance to M51, so we converted all galactic radii values for a common distance of 9.7 Mpc. The two main spiral arms are shown (green and blue). CO data are shown as triangles, HII data are shown as circles. Filled symbols (circles and triangles) are for the blue arm; unfilled symbols are shown for the green arm. The full pitch angle range (from -90° to + 90°) is shown. Most values are found in the narrow pitch angle range between 0° and 30°. The running mean pitch angle appears flat with increasing galactic radius, while local deviations reaching 20° can be seen.

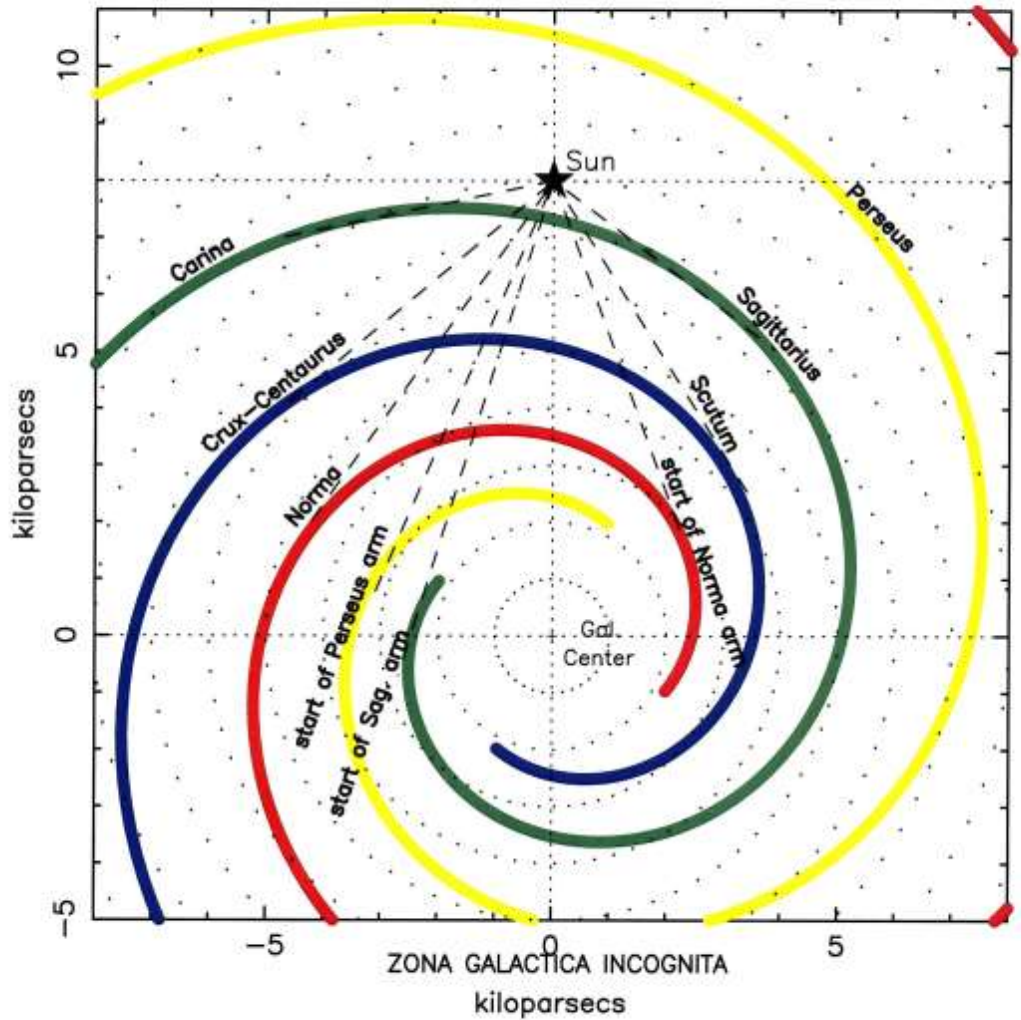

Figure 2. Newest model with a 4-arm logarithmic shape, fitted to the observed arm tangents (Vallée 2014c) and incorporating a pitch of -13.1° (Vallée 2015), over the range of 4 to 8 kpc. Arms are continued inward to about 2 kpc. The 4 spiral arms are shown (yellow Perseus, green Sagittarius, blue Scutum, red Norma), as computed here. Below the Galactic Center, the poorly known region is the so-called 'Zona Galactica Incognita'.

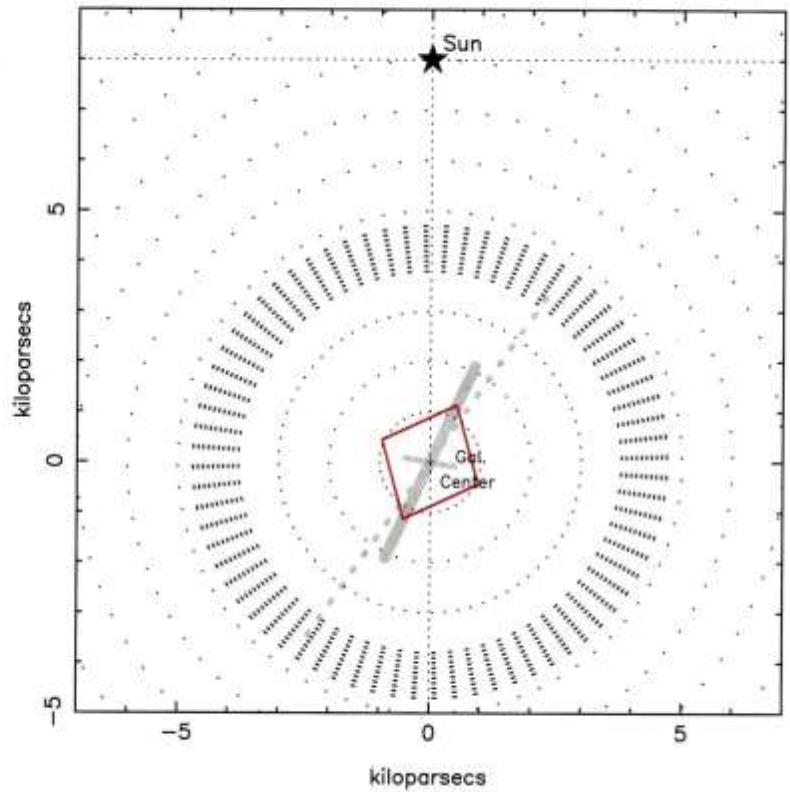

Figure 3. Map of observed elongations in the central portion of the Milky Way. A schematic of the nuclear bar (gray), long bar (dotted gray), bulge bar (thick gray) in the central zone, and of the central bulge (shown as a red diamond). The blurred molecular ring around 4 kpc is shown (radial dashes). Circles are shown every kpc from the Galactic Center (dots). The Sun is at the top (black star at top). All data scaled to a Sun-to-Galactic center distance of 8.0 kpc. Data from Tables 2 and 3.

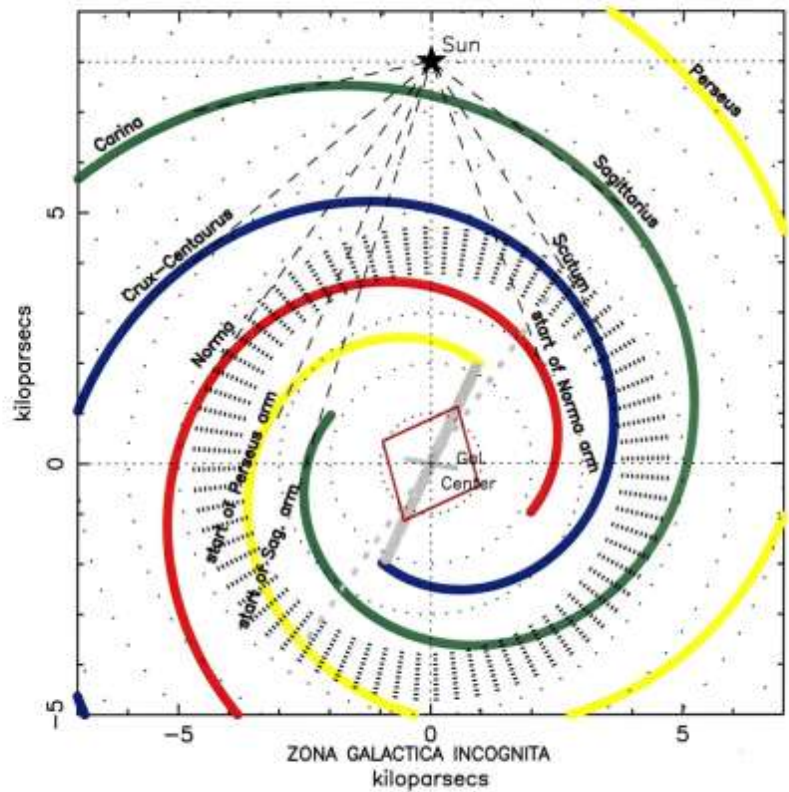

Figure 4. An overlay of the best arm model (Fig.2) with the observed objects in the inner Galaxy (fig.3).

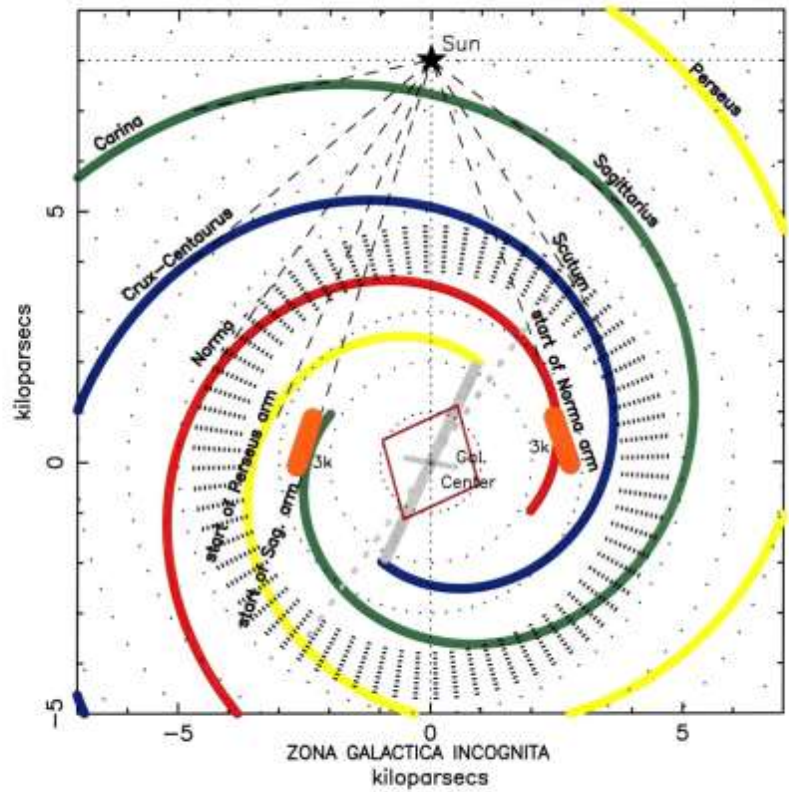

Figure 5. Map of the disk of the Galaxy. Here we added the mean longitudes of the so-called '3-kpc-arm' features (orange segments). Data from table 5.

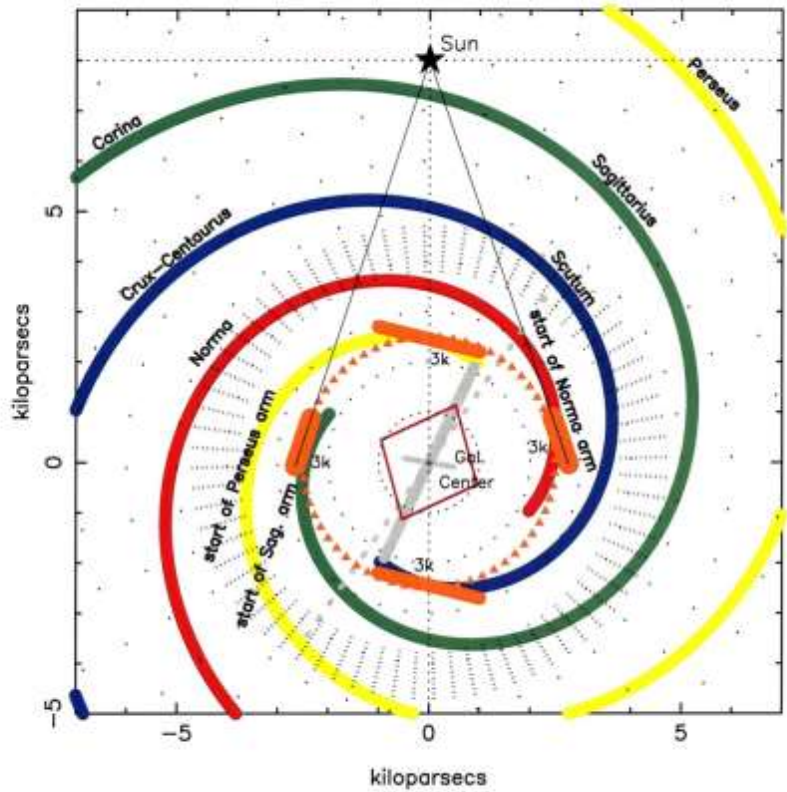

Figure 6. Composite map of the central portion of the Galaxy. We have drawn a model of the four so-called '3-kpc-arm' features (orange segments), taken as being symmetric for each of the start of a spiral arm. Some gas may be in a pseudo ring (orange triangles) at a galactic radius near 2.5 kpc.